\documentclass[12pt]{article}
\usepackage[T1]{fontenc}
\usepackage{graphicx,authblk,url}
\pdfoutput=1

\author[1]{Brandon Chrisman}
\author[1]{Allison Rabe}
\author[2]{Ranelle Ivens}
\author[2]{Sarah Lopez}
\author[2]{Alexey Shipunov}

\affil[1]{Minot High School, Minot, ND 58701}
\affil[2]{Minot State University, Minot ND, 58707}

\title{The ecological impact of flooding:\\ a study of tree damage}

\date{}

\newenvironment{Hang}{\leftskip2em\parindent-2em\parskip1ex\mbox{}}{}

\begin{document}

\maketitle

\begin{abstract}

The objective of this research was to identify factors affecting tree damage in the historical Minot flood of 2011. We hypothesized that tree height, identity, origin, and maximum water height affect in the severity of damage sustained by a tree in a flood event. All these factors were significant but highly interactive. The results from this research can influence planting practices in valleys and other flood prone areas to mitigate future damage.

\end{abstract}

\section{Introduction}

Various effects of flooding can drastically alter an ecosystem. Some of the factors that could potentially harm trees include alteration of soil factors, disturbance of plant life, and degeneration of habitat. During a flood, soil is altered by erosion, deoxygenation, and reconfiguration (Kozlowski 1997; Lake 2011). Erosion is caused by the continuous circular and turbulent motion of the water. This motion carries particles of the soil away, resulting in a loss of topsoil, loss of nutrients, and an exposure of plant roots; all three of these factors severely harm plants (Shafroth et al. 2000). The topsoil is reconfigured with the finer particles that were carried with the flood, such as sand and silt, which may to compact into finer arrangements. This compacted soil restricts the ability to hold gases such as oxygen, causing deoxygenation of the soil (Parrett 1964).

Flooding disturbs not only the abiotic factors of the environment, but also the biotic factors. Plants suffocate and decompose due to the flood waters (Parolin \& Wittmann 2010). This can result in stunting, damaging, or even killing of trees (Joly \& Crawford 1982). The warm, stagnant water is a suitable habitat for several bacteria and fungi. This can cause tree rot as the bacteria and fungi in the water inhibit the trees' natural ability to carry nutrients and water (Coder 1994). Because the part of a tree is submerged, the tree is likely to undergo suffocation if it is submerged for a long period of time. The suffocation of trees is caused by the lack of oxygen in the stagnant flood waters. If there is a faster flowing current, the water can also damage a tree's structure by breaking off branches or uprooting it altogether.

Considering both abiotic and biotic factors, flooding substantially distorts the original environment. Damage to native plant life during the flood and the change in soil configuration after the flood may provide a more suitable environment for foreign species (Yanosky 1982). Since the postdiluvian soil determines the species of plants present, a flood can restructure an ecosystem (Bratkovich 1993).

Minot, North Dakota, is positioned in the Souris River Valley. The Souris River originates in Canada, flows south into the United States at North Dakota, passes through Minot, and then turns north again to re-enter Canada. Flooding of varying severity is common along the Souris River. Despite numerous previous floods, none were as large as the flood of 2011. Record water levels from the 1969 flood were surpassed as water rose to 457.8 m above sea level. This record flood event submerged houses and crested bridges. Warm, stagnant, and hypoxic waters from summer rains covered the valley for approximately 23 days in June and July (``Timeline...''). The river banks are presently wider and contain more silt than the antediluvian banks; this was caused by erosion of the banks. The flood has also affected the plant life present in the flood plain. Vast areas of the Minot flood plain were drastically over-saturated during the flood, causing a degeneration of suitable habitats for most indigenous plant species (``Timeline...'').

Studying tree species in Minot has allowed for better understanding of the extent of damage suffered by trees during floods. By determining which trees are best adapted to survive in flooding situations, informed decisions can be made when planting trees in areas prone to flooding. This knowledge will allow for the least sustained damage of habitats after future floods.

Here, we analyze data collected regarding the impact of a flood event on tree damage. We hypothesize that tree height, genus, and maximum water level are all determining factors in the severity of damage sustained by a tree in a flood event.

\section{Materials and Methods}

We used the Minot flood plain for research. We sampled trees across the Minot flood plain in September 2011 and in Oak Park in July 2014 with tools such as clinometers, logger's tape, and a GPS device. For each tree we recorded height, water level, identity, and damage. Damage was visually accessed on a scale of 0 to 5: 0 = no damage, 1 $\le$ 25\% of the tree was damaged, 2 = 25--50\% of the tree was damaged, 3 = 50--75\% of the tree was damaged, 4 = 75--90\% of the tree was damaged, and 5 = 100\% of the tree was damaged.

We measured tree heights with a clinometer and recorded to nearest meter. Maximum water level was measured using the still visible line of water on objects surrounding the tree, such as houses or fences. We identified (Herman \& Chaput 2003) to the level of genera and measured 288 trees.

The second part of the research explored the relationship between the maximum water level and tree damage. We made a virtual transect along the 8th Street (which comes almost perpendicular from the river to the undamaged part of city) and use apple trees (\emph{Pyrus}) as a standard indicator of damage. We recorded the damage and distance from the river in meters of \emph{Pyrus} trees along this route.

In the Oak Park section, the exact GPS coordinates of trees were also recorded. This measurement will allow for accurate and consistent reevaluation over time. The damage rating has been recorded annually. 

We used the R statistical environment (R Core Team 2014) for all calculations and plots. Data categories were analyzed in various combinations in order to thoroughly evaluate relationships (Dalgaard, 2008). We analyzed the relationship between height and genera, water level and genera, tree height and water level, tree height and damage, water level and damage, genera and damage, conifers and damage, angiosperms and damage, and damage and time. Relationships between the data were all determined after inputting the data. We analyzed the data through these various combinations in order to fully understand the relationships between the factors.

All data files and R script are available in the open data repository at \url{http://ashipunov.info/shipunov/open/ecol_impact_flooding.zip}

\section{Results}

288 studied trees belonged to 12 different genera; the majority being \emph{Fraxinus} and \emph{Ulmus}. Damages recorded were mostly below 3; less than 40\% trees were damaged >50\%.

We measured interactions between the data recorded. Both height of tree and maximum water level were significantly correlated with the taxonomy (Kruskall-Wallis tests p-values $\ll 0.05$ in both cases). Tree height and water level were also weakly but significantly correlated (Spearman $\rho = 0.282$, p $\ll 0.05$). Consequently, on next steps of the data analysis we focused on the interaction between factors.

The linear models of damage included two continuous (tree height and water level) and one categorical variable (tree genus), and were different with and without interactions. With interactions, only one term (\emph{Quercus} genus) was significant whereas the model without interactions had multiple significant terms, namely tree height, water level, and \emph{Prunus}, \emph{Pyrus} and \emph{Quercus} genera (all p-values < 0.05). \emph{Prunus} and \emph{Pyrus} sustained damage more than other trees (median damage 3 while the median damage for the all data was 2), but \emph{Quercus} was damaged on the median level. 

Both linear models were significant (Fisher tests p-values $\ll 0.05$), but the adjusted R$^2$ in both cases was low (0.167 and 0.111, respectively). The second model was more optimal (AIC $= 1027.387$ vs. AIC $= 1036.728$ for the first model with interactions). Attempts to remove any terms in order to simplify the second model resulted in an increased AIC.

The damage was weakly but significantly negatively correlated with the tree height (Spearman $\rho = -0.285$, p-value $\ll 0.05$) but not with water level (Spearman $\rho = 0.022$, p-value $= 0.703$).

Post hoc pairwise comparisons of damage level between different tree genera did not reveal any significant patterns (Nemenyi test p-values $>> 0.05$), only damages to \emph{Pyrus} and \emph{Populus} were significantly different (Nemenyi test p-value $= 0.045$). 

Various contrasts were applied to our set of tree species. While native and non-native trees were not significantly different in the level of damage (Wilcoxon test p-value $= 0.452$), conifers were damaged significantly more than angiosperm trees (Wilcoxon test p-value $= 0.033$). However, without \emph{Juniperus} (i.e., only for pine family, Pinaceae trees) this significant pattern disappeared (Wilcoxon test p-value $= 0.316$). Rosaceae trees (\emph{Pyrus} and \emph{Prunus}) also suffered damage more than others (Wilcoxon test p-value $= 0.003$) but this could have been influenced by their significantly smaller sizes (Wilcoxon test p-value $\ll 0.05$).

To understand how damage is related with water level, we used transect data less influenced by interactions since it focused on only one genus (\emph{Pyrus}) and location. The linear model for the damage vs. maximum level of water on the transect returned all terms significant, adjusted R$^2$ $= 0.551$ and Fisher test p-value $\ll 0.05$.

While there was a decreasing trend of the tree damage in Oak Park (from 3.0 in 2011 to 2.0 two years later), this pattern was not significant (Friedman test p-value $= 0.905$).

\section{Discussion}

The patterns observed suggest several conclusions. First of all, the assumption of a greater tolerance to flooding of native species than nonnative species is not supported by the data. Others have observed a relationship between indigenous species and damage resistance in response to locally common disasters (Zamora-Arroyo et al. 2001), but this was not supported by our flood data. This is probably because at least some of the foreign species originated from areas with similar flooding occurrences as the Minot flood plain. 

A significant inverse relationship between tree height and damage sustained was found. While tree height contributes greatly to the amount of damage trees sustain after a flood, the data also shows the separate relationships between the taxonomy and damage. The result that certain genera show greater resistance to flooding than other genera is consistent with previous research (Parolin \& Wittmann 2010; Kozlowski et al., 2015). Conifers (Pinopsida) and \emph{Pyrus} and \emph{Prunus} (both of Rosaceae family) sustained significantly more damage than other genera. These results allow for educated decisions in the future when planting trees inside a flood zone.

The factors measured were highly interactive in determining flood damage. Water level, tree height, and genus all had significant effects on tree damage; the removal of any of these terms from the model decreased optimality. While it appears that a combination of all factors determines damage, tree height seems to exhibit the greatest influence.

The study of damage to Oak Park trees over time is still in its early stages. As it stands, the data suggest promising regeneration rates for most of the trees involved. Continued research will be conducted on these trees to gain further insight regarding tree regeneration. Drawing upon similar research, we predict that native trees will show a greater rate of regeneration than nonnative trees (Zamora-Arroyo et al. 2001).

\section{Acknowledgements}

Thank you to Joe Super (Minot High School) for providing equipment and motivation. Many thanks to the students of Minot State University Biology 154 class for their contributions in data collection. We are grateful to reviewers for their valuable suggestions and corrections.

\section{Literature cited}

\begin{Hang}

Bratkovich, S, L. Burban, S. Katovich, C. Locey, J. Pokorny, and R. Wiest. 1993. Flooding and its Effect on Trees. Misc. Publ. Newtown Square, PA: U.S. Dept. of Agriculture, Forest Service, Northern Area State \& Private Forestry.

Coder, K. D. 1994. Flood Damage to Trees. University of Georgia School of Forest Resources Extension, publication FOR 94-061. \url{http://warnell.forestry.uga.edu/service/library/index.php3?docID=104}. Acces\-sed 22 November 2014.

Dalgaard, P. 2008. Introductory statistics with R. Springer Science \& Business Media, New York.

Parolin, P., and F. Wittmann. 2010. Struggle in the flood: tree responses to flooding stress in four tropical floodplain systems. AoB PLANTS 2010: plq003.

Herman, D. E., and L. J. Chaput. 2003. Trees and shrubs of North Dakota. NDSU, Fargo.

Joly, C. A., and R. M. M Crawford. 1982. Variation in Tolerance and Metabolic Responses to Flooding in Some Tropical Trees. Journal of Experimental Botany. 33:799--809.

Kozlowski, G., Stoffel, M., Betrisey, S., Cardinaux, L., and Mota, M. 2015. Hydrophobia of gymnosperms: myth or reality? A global analysis. Ecohydrology. 8:105--112.

Kozlowski, T. T. 1997. Responses of woody plants to flooding and salinity. Tree Physiology Monograph. 1:1–29.

Lake, P. S. 2011. Drought and Aquatic Ecosystems: Effects and Responses. Chichester, West Sussex. : Wiley Blackwell.

Parrett, A. 1964. Montana's Worst Natural Disaster. Montana: The Magazine of Western History 2:20.

R Core Team. 2014. R; A language and environment for statistical computing. R Foundation for Statistical Computing, Vienna, Austria.

Shafroth, P. B., J .C. Stromberg, and D. T. Patten. 2000. Woody riparian vegetation response to different alluvial water table regimes. Western North American Naturalist, 60:66--76.

Timeline of Minot's 2011 Flood. 2011 August 07. Minot (ND): Minot Daily News. \url{http://www.minotdailynews.com/page/content.detail/id/ 557475/Timeline-of-Minot-s-2011-flood.html?nav=5010}. Acces\-sed 16 March 2015.

Yanosky, T. M. 1982. Effects of Flooding Upon Woody Vegetation Along Parts of the Potomac River Flood Plain. Geological Survey Professional Paper 1206, Distr. Branch USGS.

Zamora-Arroyo, F., P. Nagler, M. Briggs, D. Radtke, H. Rodriquez, J. Garcia, C. Valdes, A. Huete, and E. Glenn. 2001. Regeneration of native trees in response to flood releases from the United States into the delta of the Colorado River, Mexico. Journal of Arid Environments 49:49--64

\end{Hang}

\section{Figures}

\begin{figure}[ht]\centering

\includegraphics[width=.9\textwidth]{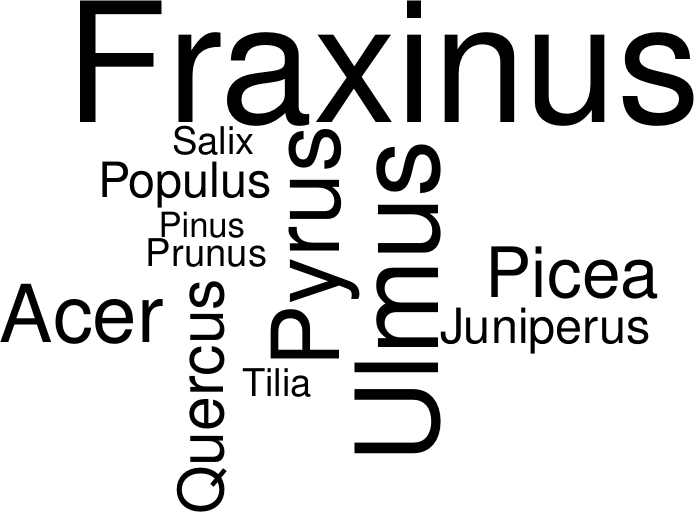}

\caption{Generic content of the Minot flooded tree population are shown in this figure. The text size is proportional to the prevalence of the particular genus. Only genera that had more than 4 trees in the data were included in this figure. The data for this figure was collected on September 2011 in the flooded are of Minot, North Dakota, USA.}\label{1}

\end{figure}

% ===

\begin{figure}[ht]\centering

\includegraphics[width=\textwidth]{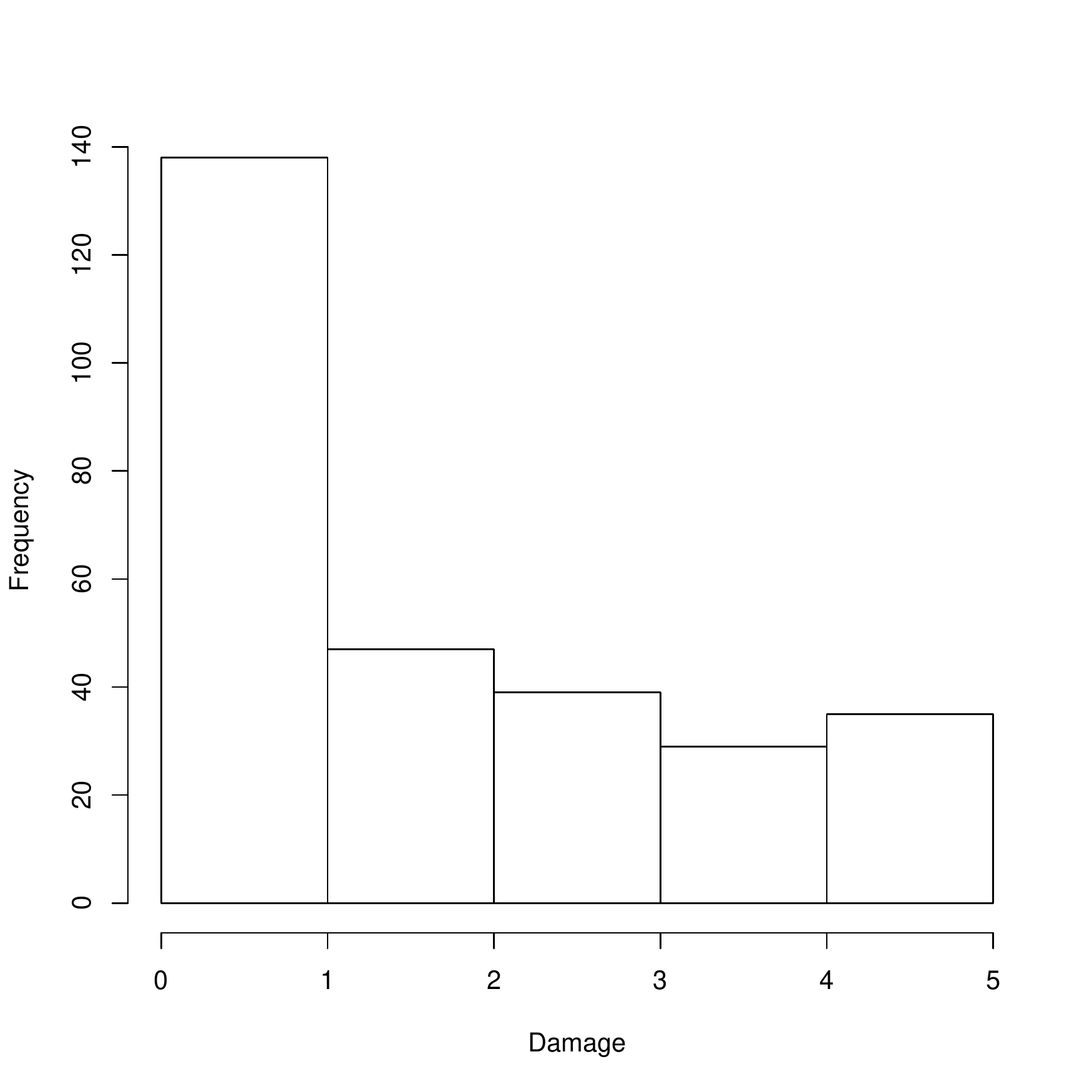}

\caption{Histogram of the tree damage. This figure shows the dispersal of trees among the different damage groups. All genera were included in this test and figure. The data for this figure was collected on September 2011 in the flooded area of Minot, North Dakota, USA.}\label{2}

\end{figure}

% ===

\begin{figure}[ht]\centering

\includegraphics[width=\textwidth]{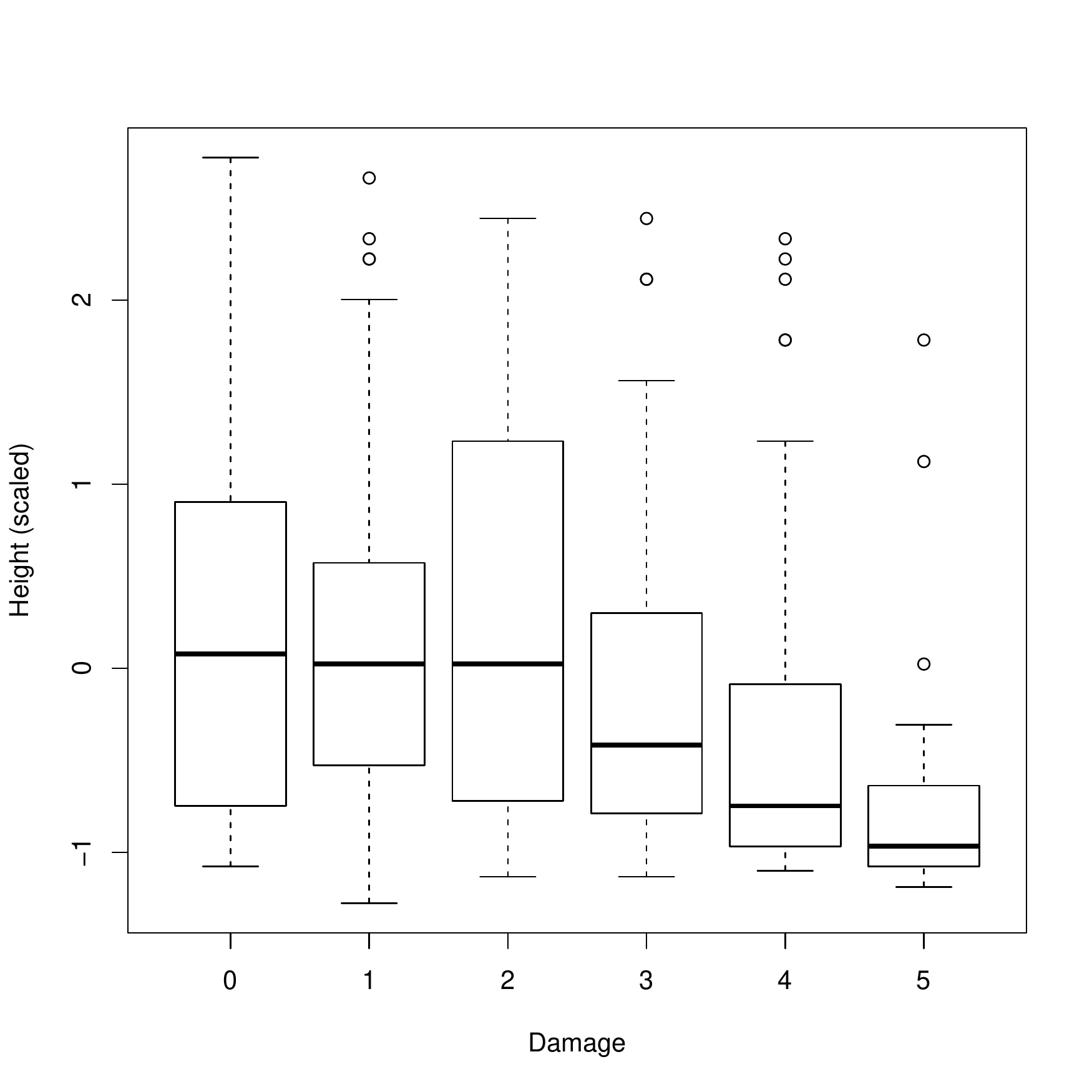}

\caption{For every damage level, tree height boxplot is shown. The higher levels of damage correspond with the significantly smaller tree sizes. All genera were included in this test and figure. The data for Figure 3 was collected on September 2011 in the flooded area of Minot, North Dakota, USA.}\label{3}

\end{figure}

% ===

\begin{figure}[ht]\centering

\includegraphics[width=\textwidth]{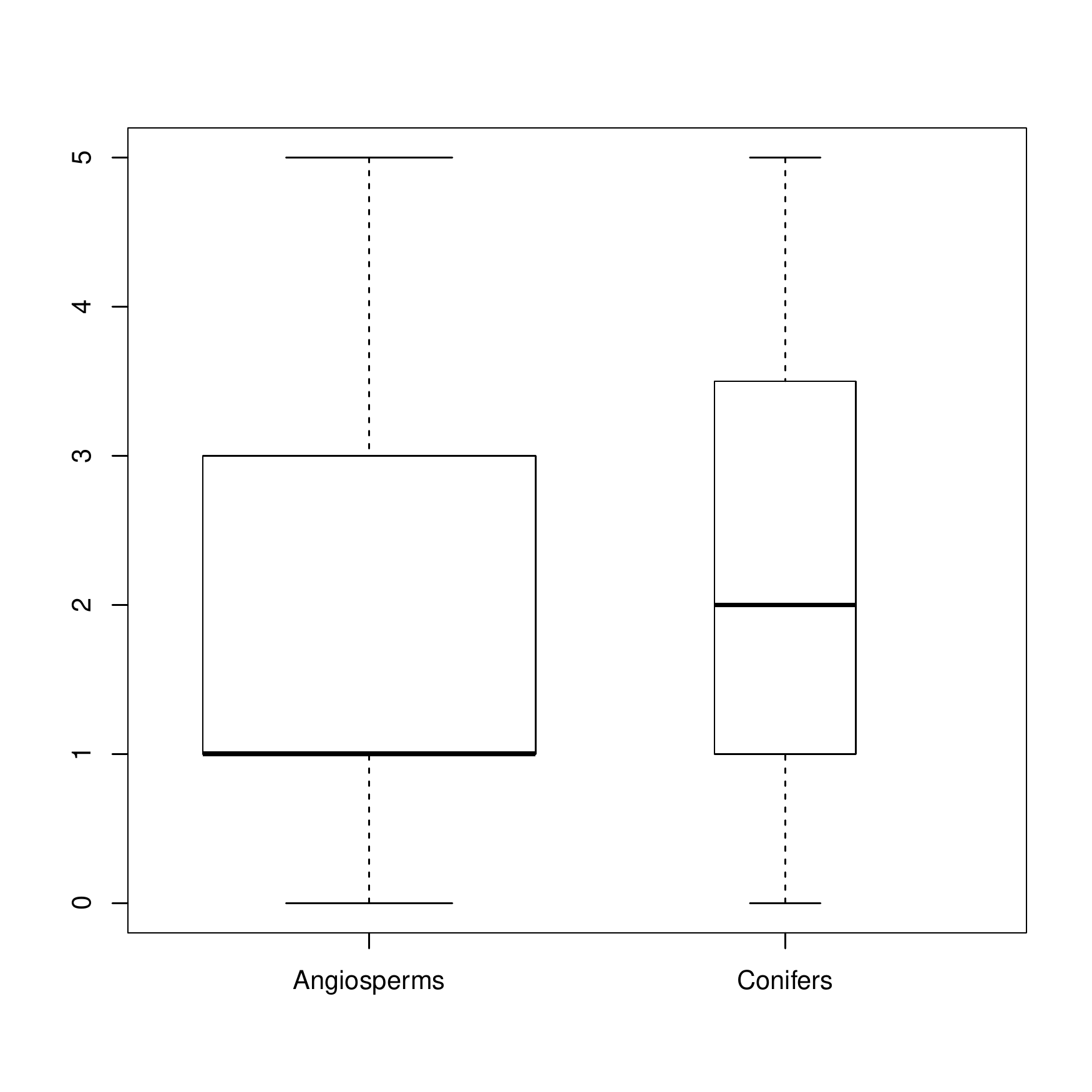}

\caption{Boxplots showing damage in conifers and non-conifers (angiosperms). All collected genera were included in this test and figure after being divided into the correct category. The data for this figure was collected on September 2011 in the flooded area of Minot, North Dakota, USA.}\label{4}

\end{figure}

% ===

\begin{figure}[ht]\centering

\includegraphics[width=\textwidth]{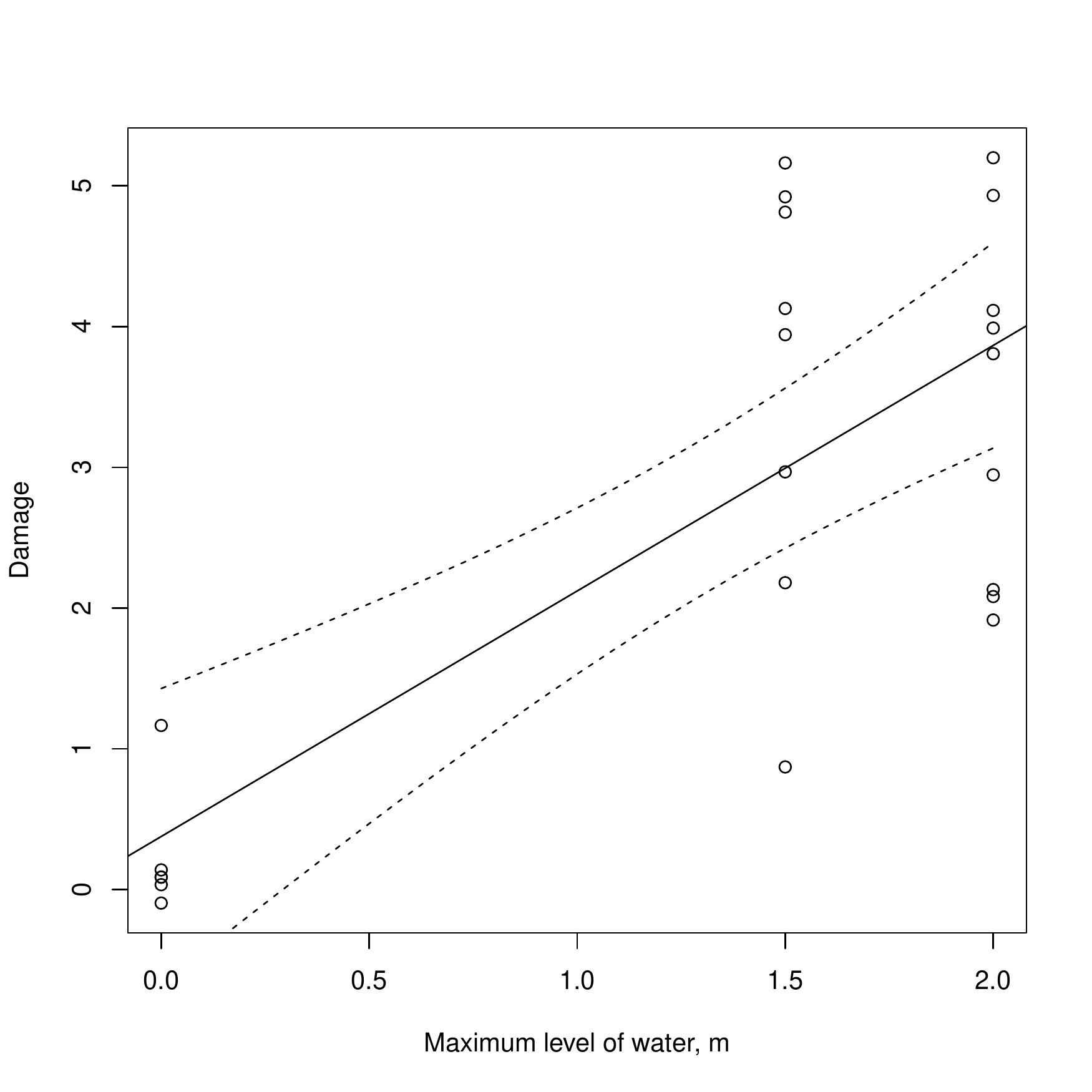}

\caption{Correspondence between the maximum level of water on the transect, and damage of apple trees (\emph{Pyrus}). Dotted lines show the 95\% confidence interval. Only Pyrus representatives were included for this test and figure. The data for this figure was collected on September 2011 along 8th Street in Minot, North Dakota, USA.}\label{5}

\end{figure}

\end{document}